\documentclass[12pt]{article} 

\renewcommand{\thefootnote}{\fnsymbol{footnote}}
\usepackage{amsbsy,amssymb,latexsym,amsfonts,amsmath}
\usepackage{mathrsfs}
\usepackage{graphicx}
\usepackage{cite}
\usepackage{bm} 
\numberwithin{equation}{section}
\newcommand{\bel}[1]{\begin{equation}\label{#1}}                     
\newcommand{\bal}[1]{\begin{eqnarray}\label{#1}}                     
\newcommand{\be}{\begin{equation}}
\newcommand{\ee}{\end{equation}}
\newcommand{\im}{\mathrm{i}}
\newcommand{\ex}{\mathrm{e}}
\newcommand{\de}{\mathrm{d}}

\newcommand{\ul}[1]{\underline{#1}}
\begin{document}
%%%%%%%%%%%%%%%%%%%%%%%%%%%%%%%%%%%%%%%%%%%%%%%%%%%%%
%%%%%%%%%%%%%%%%%%%%%%%%%%%%%%%%%%%%%%%%%%%%%%%%%%%%%
%
% title page
%
%%%%%%%%%%%%%%%%%%%%%%%%%%%%%%%%%%%%%%%%%%%%%%%%%%%%%
\begin{titlepage}
%%%%%%%%%%%%%%%%%%%%%%%%%%%%%%%%%%%%%%%%%%%%%%%%%%%%%
\begin{flushright}
\normalsize
%\filename
~~~~
NITEP 196\\
February, 2024 \\
\end{flushright}
%%%%%%%%%%%%%%%%%%%%%%%%%%%%%%%%%%%%%%%%%%%%%%%%%%%%%

\vspace{15pt}

%%%%%%%%%%%%%%%%%%%%%%%%%%%%%%%%%%%%%%%%%%%%%%%%%%%%%
\begin{center}
{\LARGE Large order behavior near  the AD point:} \\
{\LARGE the case of $\mathcal{N} =2$, $su(2)$, $N_f =2$}
\end{center}
%%%%%%%%%%%%%%%%%%%%%%%%%%%%%%%%%%%%%%%%%%%%%%%%%%%%%

\vspace{23pt}

%%%%%%%%%%%%%%%%%%%%%%%%%%%%%%%%%%%%%%%%%%%%%%%%%%%%%
\begin{center}
{ Chuan-Tsung Chan$^a$\footnote{e-mail: ctchan@go.thu.edu.tw}, 
H. Itoyama$^{b,c}$\footnote{e-mail: itoyama@omu.ac.jp},
  and  R. Yoshioka$^{b,c}$\footnote{e-mail: ryoshioka@omu.ac.jp}  }\\
%%%%%%%%%%%%%%%%%%%%%%%%%%%%%%%%%%%%%%%%%%%%%%%%%%%%%

\vspace{18pt}

%%%%%%%%%%%%%%%%%%%%%%%%%%%%%%%%%%%%%%%%%%%%%%%%%%%%%

$^a$ \it Department of Applied Physics, Tunghai University, \\ 
Taichung, 40704, Taiwan \\
\vspace{5pt}

$^b$ \it Nambu Yoichiro Institute of Theoretical and Experimental Physics (NITEP), Osaka Metropolitan University \\ 

$^c$\it Osaka Central Advanced Mathematical Institute (OCAMI), Osaka Metropolitan University\\

3-3-138, Sugimoto, Sumiyoshi-ku, Osaka, 558-8585, Japan \\

\end{center}
%%%%%%%%%%%%%%%%%%%%%%%%%%%%%%%%%%%%%%%%%%%%%%%%%%%%%

\vspace{20pt}

\begin{center}
Abstract\\
\end{center}
%%%%%%%%%%%%%%%%%%%%%%%%%%%%%%%%%%%%%%%%%%%%%%%%%%%%%
A non-perturbative effect in $\kappa$ (renormalized string coupling) 
 obtained from the large order behavior in the vicinity of the prototypical Argyres-Douglas critical point 
 of  $su(2)$, $N_f =2$, $\mathcal{N} =2$ susy gauge theory can be studied  in the GWW unitary matrix model
with the log term: the one as the work done against the barrier of the effective potential 
 by a single eigenvalue lifted from the sea
and the other as a non-perturbative function contained in the solutions of the nonlinear differential equation PII  that goes beyond the asymptotic series.
 The leading behaviors are of the form $\exp (-\frac{4}{3}\frac{1}{\kappa} \, (1, \left(\frac{s}{K}\right)^{\frac{3}{2}} ))$ respectively.  
 We make comments on their agreement.

%%%%%%%%%%%%%%%%%%%%%%%%%%%%%%%%%%%%%%%%%%%%%%%%%%%%%

\vfill

\end{titlepage}

%%%%%%%%%%%%%%%%%%%%
\renewcommand{\thefootnote}{\arabic{footnote}}
\setcounter{footnote}{0}
%%%%%%%%%%%%%%%%%%%%

%%%%%%%%%%%%%%%%%%%%%%%%%%%%%%%%%%%%%%%%%%%%%%%%%%%%%
%%%%%%%%%%%%%%%%%%%%%%%%%%%%%%%%%%%%%%%%%%%%%%%%%%%%%
\section{Introduction}
%%%%%%%%%%%%%%%%%%%%%%%%%%%%%%%%%%%%%%%%%%%%%%%%%%%%%
%%%%%%%%%%%%%%%%%%%%%%%%%%%%%%%%%%%%%%%%%%%%%%%%%%%%%

Instanton expansion has continued to play a basic role in the development of
supersymmetric gauge theory ever since the exact determination of low energy effective
action (LEEA) of $\mathcal{N}=2$ SUSY gauge theory in four spacetime dimensions  \cite{SW9407,SW9408}. 
String theoretic corrections have been put in accord with mathematics which enables a proper instanton counting, and which avoids  volume divergence \cite{Nekr0206,Naka1999,NY0306}.
The $\beta$-deformed matrix models of eigenvalue type are seen to generate all these machineries in a concise manner \cite{DV0909,IMO0911} 
as the partition functions in general are identified with integral representation  \cite{DF1984,MMS1001,IO1003} of the regular and irregular $2d$ conformal blocks,  
which in turn successfully generate \cite{IO1003,IOYone1008} 
the $4d$ instanton expansion \cite{AGT0906}. 
In this letter, we focus upon another well-known expansion, namely, that in $1/N$ in the matrix model side which translates into that in $g_s$ for fixed flavor masses (eq.\eqref{dictionary}). 
We will search for a non-perturbative effect associated at $\beta=1$ which is much less explored in gauge theory while, in the theory of $2d$ gravity, this has been a central theme \cite{BK1990,GM1990NPB,GM1990PRL,DS1990}. 
A key ingredient is the presence of a critical singularity in the large $N$ limit and the attendant double scaling limit that enhances the contributions from higher orders in $1/N$.\footnote{For non-perturbative effects other than this one, see, for example, \cite{MMM0909Z,IYone1104}}

In a series of papers \cite{IOYone1008,IOYano1805,IOYano1812,IOYano2019,IOYano1909,IOYano2103}, 
we obtained, among other things, the one-matrix model for irregular $2d$ conformal block that corresponds to $4d$ $\mathcal{N}=2$, $N_f =2$, $su(2)$ instanton partition function. 
See also \cite{Gaio0908,MMM0909O,BMT1112,GT1203,NR1207}. 
In obtaining this from $2d$ conformal block for $N_f=4$ by a limiting procedure with contour deformation involved, we observed that the model naturally turned into a unitary matrix model of GWW type \cite{GW1980,Wadi1980,Wadi1212} 
with the addition of the log potential. 
The notion of the double scaling limit is effective here also as the moduli space of vacua is known to contain an IR fixed point of Argyress-Douglas type \cite{AD9505,APSW9511,KY9712,Xie1204}. 
In the case of $\beta=1$, 
we have been successful in deriving the Painlev\'e II (PII) equation with an accessory parameter in the double scaling limit from this matrix model, using the method of orthogonal polynomials followed by a set of recursion relations (string eqs.). 
We find it necessary to work on the partition function without specifying the filling fraction, namely, the Coulomb moduli $a$.
This one is energetically favored and the  preference for this alternative over the instanton partition function has been 
the common practice known in the study at $N_f=4$ \cite{GIL1207,MM1707} that predates the current one. 

The above procedure serves as the prototypical study of the vicinity of the AD point,
using the matrix model. In the higher rank case where the multi-matrix model of quiver type
is relevant \cite{IMO0911,MMM1991}, 
we have resorted to the symmetry argument to identify  the AD point \cite{IOYosh2210,IOYosh2212}. 
The ``unitarization" made by the same limiting procedure as in the $su(2)$ case has been crucial to make the point of maximal symmetry in flavor mass moduli manifest.

In this paper, we continue our prototypical study at $N_f=2$.
The equation PII itself can not only generate $1/N$ expansion in the scaling limit as an asymptotic one but also contains a non-perturbative piece on top of it.  
This piece can be exhibited for large scaling variable denoted by $s$ in this letter by a simple analysis  in what follows.
After giving a minimal account of formulae for this introduction in section two (leaving the derivations to the original papers), 
we extract from PII for the susceptibility function $u(s)$ a non-perturbative piece contained beyond the asymptotic series in section three. 
This is a well-known WKB analysis applied to PII. 
We pay a particular attention to the parameter dependence of the scaling variable and 
 that of the scaling function.
We make a brief digression on the dependency on parameters 
 associated with the canonical transformation from $(u, p_u)$ to $(y, p_y)$, 
which puts PII in the standard form in mathematics literature \cite{Okam2009}. 

 In matrix models, the double scaling limit is taken in the vicinity of the singularity associated with the maximum of a potential. In section four, we study the work
 done against the barrier of the effective potential by a single eigenvalue 
 lifted from the sea of the filled ones in the continuum limit.
A closely related object is the vacuum to vacuum tunneling fluctuation where a single eigenvalue hops from one edge of the eigenvalue distribution to the other through the forbidden zone, namely outside the cut. 
    
 In section five, we compare the result on the non-perturbative piece in $u(s)$ 
 in section three and that of section four.
 For that to be possible, this non-perturbative function must be evaluated at a particular value.
 We see that the evaluation at the point which is actually the end point of the free energy integral
 in  the double scaling limit makes the two objects agree.

%%%%%%%%%%%%%%%%%%%%%%%%%%%%%%%%%%%%%%%%%%%%%%%%%%%%%
%%%%%%%%%%%%%%%%%%%%%%%%%%%%%%%%%%%%%%%%%%%%%%%%%%%%%
\section{Some preliminaries} 
%%%%%%%%%%%%%%%%%%%%%%%%%%%%%%%%%%%%%%%%%%%%%%%%%%%%%
%%%%%%%%%%%%%%%%%%%%%%%%%%%%%%%%%%%%%%%%%%%%%%%%%%%%%
As is stated in the introduction, the model which is relevant to our problem is the unitary matrix model
\begin{equation}
Z_{U(N)} = \frac{1}{\mathrm{Vol}(N)} \int [\de U] \exp\Big(
 \mathrm{Tr}\, W_U(U) 
\Big), 
\end{equation}
with the potential\footnote{For an extension of the model and related aspects, see, for example, \cite{PS1990PRL,PS1990NPB,Mina9111,Hisa9611,AMM1610,CIO9305,IOSY1609,MOT1909,Oota2112}.}
\begin{equation}\label{W_U}
W_U(w) = -q_{02} \left(w + \frac{1}{w} \right) + M \log w. 
\end{equation}
The relation of the parameters $N$, $M$ and $q_{02}$ with the 4d parameters are listed below.

%%%%%%%%%%%%%%%%%%%%%%%%%%%%%%%%%%%%%%%%%%%%%%%%%%%%%
\subsection{0d-4d connection \rm\small \cite{IO1003,IOYone1008}}
%%%%%%%%%%%%%%%%%%%%%%%%%%%%%%%%%%%%%%%%%%%%%%%%%%%%%
In the $\beta$ deformed matrix model for the $N_f=2$, 
there are four net 0d parameters aside from $q_{02}$,  
\begin{equation}
\alpha_{1+2},~~ \alpha_{3+4},~~ \beta,~~ 
N_L,~~N_R,~~(N \equiv N_L+N_R) , 
\end{equation}
under the constraint: 
\begin{equation}
\alpha_{1+2} + \alpha_{3+4} + 2\sqrt{\beta} N = 2 Q_E,~~~~~
c = 1-6 Q_E^2,~~ Q_E = \sqrt{\beta} - \frac{1}{\beta}. 
\end{equation}
The number of free parameters reduces to two by setting $\beta=1$ and just referring to $N$. 

Likewise, there are four 4d parameters aside from $\Lambda_2$, the 4d scale parameter: 
\begin{equation}
\frac{\epsilon_1}{g_s},~ \frac{a}{g_s},~ \frac{m_1}{g_s},~ \frac{m_2}{g_s},~~~~~
\epsilon_1 = \sqrt{\beta} g_s,~~ \epsilon_2 = -\frac{g_s}{\sqrt{\beta}},~~ \epsilon = \epsilon_1+\epsilon_2.
\end{equation}
The number reduces to two by setting $\epsilon_1/g_s = 1$ and not referring to $a$. 

These parameters are related by the 0d-4d dictionary: 
\begin{equation}\label{dictionary}
\begin{split}
&\alpha_{1+2} = \frac{1}{g_s} \left( 2 m_2 + \epsilon \right) ,\\
&\alpha_{3+4} = \frac{1}{g_s} \left( 2m_1 + \epsilon \right), \\
& 
N 
= \frac{a-m_2}{g_s} - \frac{a+m_1}{g_s}
= - \frac{m_1+m_2}{g_s}.
\end{split}
\end{equation}

Separately, the parameter for the 4d instanton expansion reads   
\begin{equation}
q_{02} = \frac{\Lambda_2}{2g_s} \equiv \frac{1}{2}\frac{1}{\underline{g}_s}. 
\end{equation}
and $M$ in eq.  \eqref{W_U} is related to the 4d parameter by 
\begin{equation}\label{M}
M \equiv \alpha_{3+4} + N = \frac{m_1-m_2}{g_s}. 
\end{equation}
We have seen in eq. \eqref{dictionary} and eq. \eqref{M} that, by unitarization, 4d meaning of the two independent 0d parameters $N$, $M$ has become manifest. 

%%%%%%%%%%%%%%%%%%%%%%%%%%%%%%%%%%%%%%%%%%%%%%%%%%%%%
\subsection{String equation from orthogonal polynomials \rm\small \cite{IOYano1805,IOYano1812,IOYano2019}}
%%%%%%%%%%%%%%%%%%%%%%%%%%%%%%%%%%%%%%%%%%%%%%%%%%%%%
The partition function with no specification of the filling fraction, which is denoted by $\underline{Z}_{U(N)}$,  can be expressed in terms of the  orthogonal polynomials \cite{Bess1979,IZ1980}: 
\begin{equation}
\underline{Z}_{U(N)} = h_0^N \prod_{j=1}^{N-1} ( 1 - R_j^2)^{N-j},  
\end{equation}
where 
\begin{equation} \label{p}
p_n(0) \equiv A_n = R_n D_n,~~~~~~
\tilde{p}_n(0) \equiv B_n = \frac{R_n}{D_n}, 
\end{equation}
and $h_0$ is normalization coefficient.  
The product $R_n = A_n B_n$  is going to be identified as a susceptibility.

Let us introduce $\xi_n \equiv R_n^2$, $\eta_n \equiv n \underline{g}_s$,  $\zeta \equiv M  \underline{g}_s$.
We obtain the string equation: 
\begin{align}\label{StEq}
0=& \eta_n^2 \Bigl[ 
\xi_n^2 (1 - \xi_n)^2  - \eta_n^2 \, \xi_n^2 + \zeta^2 (1 - \xi_n)^2
\Bigr] \nonumber \\
& +\frac{1}{2} \, \eta_n^2 \, \xi_n\, ( 1 - \xi_n)^2 (\xi_{n+1} - 2\, \xi_n + \xi_{n-1}) 
 -  \frac{1}{16} (1 - \xi_n)^4 ( \xi_{n+1} - \xi_{n-1})^2. 
\end{align} 

%%%%%%%%%%%%%%%%%%%%%%%%%%%%%%%%%%%%%%%%%%%%%%%%%%%%%
\subsection{The double scaling  limit \rm\small \cite{IOYano1805,IOYano1812,IOYano2019}}
%%%%%%%%%%%%%%%%%%%%%%%%%%%%%%%%%%%%%%%%%%%%%%%%%%%%%
In the planar limit, the string equation \eqref{StEq} becomes  quartic in $\xi$. 
The three out of the four roots in $\xi$ degenerate to zero at $\eta = \pm 1,~ \zeta = 0$.

Let us rewrite the potential for the sake of taking a scaling limit: 
\begin{equation}
W_N(w) = \frac{N}{\tilde{S}} \left\{ -\frac{1}{2} \left( w+ \frac{1}{w} \right) + \zeta \log w\right\}. 
\end{equation}
Here, we have chosen to regard the two parameters as
\begin{equation}
\tilde{S} \equiv N \underline{g}_s = -\frac{m_1+m_2}{\Lambda_2},  
\end{equation}
and
\begin{equation} 
\zeta \equiv M \underline{g}_s = \frac{m_1-m_2}{\Lambda_2} , 
\end{equation}
while $1/N$ is the expansion parameter. 

The double scaling limit is the limit where the fine tuning of $\tilde{S}$ (the critical scaling)  and the large $N$ limit are simultaneously arranged so as to turn the 2nd difference in eq.\eqref{StEq} into the 2nd derivative. Let 
\begin{equation}\label{dsl}
x = \frac{n}{N},~~~ \eta_n = \tilde{S} x = 1 - \frac{1}{2} a^2 s,~~~ 
\xi_n = \xi(n/N) \equiv a^2 u(s),~~~
\zeta = \frac{M\tilde{S}}{N}. 
\end{equation}
Setting $a^3 = \frac{1}{N}$, we obtain the Painlev\'{e} II equation from eq. \eqref{StEq}: 
\begin{equation}\label{PII:u}
u'' = \frac{(u')^2}{2u} + u^2 - \frac{1}{2} su - \frac{M^2}{2u}. 
\end{equation}

%%%%%%%%%%%%%%%%%%%%%%%%%%%%%%%%%%%%%%%%%%%%%%%%%%%%%
%%%%%%%%%%%%%%%%%%%%%%%%%%%%%%%%%%%%%%%%%%%%%%%%%%%%%
\section{WKB analysis to PII linearized}
%%%%%%%%%%%%%%%%%%%%%%%%%%%%%%%%%%%%%%%%%%%%%%%%%%%%%
%%%%%%%%%%%%%%%%%%%%%%%%%%%%%%%%%%%%%%%%%%%%%%%%%%%%%

%%%%%%%%%%%%%%%%%%%%%%%%%%%%%%%%%%%%%%%%%%%%%%%%%%%%%
\subsection{Non-perturbative function to $u$ asymptotic series}\label{npf:u}
%%%%%%%%%%%%%%%%%%%%%%%%%%%%%%%%%%%%%%%%%%%%%%%%%%%%%
In this subsection, 
we consider a non-perturbative effect associated with PII in the form of eq. \eqref{PII:u} 
that goes beyond the asymptotic series.\footnote{For the analysis of PI at 2d gravity, \cite{Shen1990,Davi1991,Davi9212,EZJ9310}.} 
The series begins with the planar solution: 
\begin{equation}\label{u:asym}
 u_{\rm asym} = \frac{1}{2} s + \sum_{n=1}^{\infty} u_n s^{-n+1}. 
\end{equation} 
Let $u$ and $\tilde{u}$ be the two solutions with the same asymptotics and
\begin{equation}
\ul{z} \equiv \tilde{u} - u, ~~~~~
|\ul{z}| \ll |u|,~ |\tilde{u}|. 
\end{equation}
After linearization,  we obtain
\begin{equation}\label{eq:ulz}
 \ul{z}'' =  \left( \frac{u'}{u} \right) \ul{z}' - \frac{1}{2} \frac{u'^2}{u^2} \ul{z} 
 + \left( 2u - \frac{1}{2} s + \frac{M^2}{2} \frac{1}{u^2} \right) \ul{z}. 
\end{equation}
Let the WKB ansatz be
\begin{equation}\label{WKB:ulz}
\frac{\ul{z}'}{\ul{z}} \equiv h \sqrt{u} + c (\log u)' + \cdots . 
\end{equation}
with $h$ and $c$ to be determined.
Taking a derivative of eq. \eqref{WKB:ulz}, using eqs. \eqref{eq:ulz}, \eqref{WKB:ulz} 
 and the planar solution
\begin{equation}
 u \approx \frac{1}{2} s, 
\end{equation}
 for large $s$,
we obtain, to the leading order, $h^2 =1$. In the next leading order, 
we obtain $c= \frac{1}{4}$.
Integrating \eqref{WKB:ulz} once more, we determine
the non-perturbative function up to the multiplicative factor 
\begin{equation}\label{ulz}
 \ul{z} \approx (\text{const}) \, s^{\frac{1}{4}}  \exp \left(
  -\frac{1}{\sqrt{2}} \frac{2}{3} s^{\frac{3}{2}} 
 \right). 
\end{equation}

%%%%%%%%%%%%%%%%%%%%%%%%%%%%%%%%%%%%%%%%%%%%%%%%%%%%%
\subsection{Parameter dependence of PII}
%%%%%%%%%%%%%%%%%%%%%%%%%%%%%%%%%%%%%%%%%%%%%%%%%%%%%
In order to make exploit our simple result eq. \eqref{ulz} later, 
we would like to better understand the parameter dependence of PII and 
physical quantities upon the double scaling limit.  
Let us recall that the finite $N$ free energy of the matrix model is given by 
\begin{align}
&\tilde{W}_N \equiv N^2 \log F 
= -N \log \left( I_M (1/\ul{g_s} ) \right) + W_N,\\ 
&W_N = - N^2 \sum_{n=1}^N \frac{1}{N}\left( 1- \frac{n}{N} \right) \log(1-\xi_n), 
\end{align}
where $I_M(x)$ is the modified Bessel function \cite{IOYano1812}. 
Here, we work on a more general parametrization of the double scaling limit than eq. \eqref{dsl}: 
\begin{equation}
N = a^{-3},~~~~~
\eta \equiv 1- \frac{1}{2} \alpha a^2 s,~~~~~
\xi(\eta) \equiv \beta a^2 u(s),~~~~~
\tilde{S} = 1 - \kappa^{-\frac{2}{3}} a^2.
\end{equation}
Here, we have introduced the two parameters $\alpha$ and $\beta$ as the multiplicative
factors of the scaling variable $s$ and the scaling function $u$ respectively
and $\kappa=\frac{1}{(1-\tilde{S})^{3/2}N}$ is the renormalized string coupling. Note that the interval $[0, \tilde{S}]$ for $\eta$ translates into the interval $[\infty, K]$ for $s$ with
\begin{equation}
K = \frac{2\kappa^{-\frac{2}{3}}}{\alpha}. 
\end{equation}
The free energy in the double scaling limit is given by 
\begin{equation}\label{W}
W = \lim_{N \to \infty} W_N = \frac{1}{4} \alpha^2\beta \int_K^{\infty} \de s\, (s-K) u(s). 
\end{equation}
Eq. \eqref{PII:u} of PII gets scaled accordingly: 
\begin{equation}\label{PII:u,general}
u'' = \frac{(u')^2}{2u} + \alpha^2 \beta u^2 - \frac{1}{2}\alpha^3 s u - \frac{\alpha^2 M^2}{2 \beta^2 u}.  
\end{equation}
We conclude that, for \eqref{PII:u,general}, eq. \eqref{ulz} is replaced by
\begin{equation}\label{ulz:general}
 \ul{z}(s) \approx (\text{const}) \, (\alpha s)^{\frac{1}{4}}  \exp \left(
  -\frac{1}{\sqrt{2}} \frac{2}{3} (\alpha s)^{\frac{3}{2}} 
 \right). 
\end{equation}

%%%%%%%%%%%%%%%%%%%%%%%%%%%%%%%%%%%%%%%%%%%%%%%%%%%%%
\subsection{Canonical transformation of PII}
%%%%%%%%%%%%%%%%%%%%%%%%%%%%%%%%%%%%%%%%%%%%%%%%%%%%%
We have also investigated another form of \eqref{PII:u} known in mathematics\footnote{See, for example, \cite{KMNOY0403,Okam2009}.}  
 by a canonical transformation, paying a good attention to its dependency on parameters.
While we will not make exploit this later, we include here the upshot briefly.  
There is one-parameter family of Hamiltonians $H(u, p_u)$ that derives \eqref{PII:u}: 
\begin{equation}
 H(u,p_u) = \frac{\tilde{\alpha}}{2} p_u^2u - \frac{1}{2\tilde{\alpha}} u^2 
 + \frac{1}{2\tilde{\alpha}}su - \frac{M^2}{2u\tilde{\alpha}}. 
\end{equation}

Next, we make a canonical transformation to the system which 
possesses a standard kinetic term up to coefficient. 
This is done by 
\begin{equation}
u = -\frac{1}{\xi} p_y, ~~~~~
p_u = \xi y + \eta \frac{M}{p_y}. 
\end{equation}
We obtain 
\begin{equation}\label{Hamiltonian2:y}
  \hat{H}(y,p_y) = 
  -\frac{1}{2\tilde{\alpha}\xi^2}p_y^2 -\frac{\tilde{\alpha}\xi}{2}p_y y^2 - \frac{1}{2\tilde{\alpha}\xi} s p_y - \tilde{\alpha} \eta M y.
\end{equation}
Here, we discarded the term proportional to $1/p_y$, setting 
\begin{equation}
 \eta^2 = \frac{\xi^2}{{\tilde{\alpha}}^2}. 
\end{equation}
Eq. of motion for $y$ is 
\begin{equation}\label{eom:y}
 y'' = \frac{\tilde{\alpha}^2\xi^2}{2} y \left( y^2 + \frac{s}{\tilde{\alpha}^2\xi^2} \right) -\frac{\eta}{\xi^2} M -\frac{1}{2\tilde{\alpha}\xi}.
\end{equation}
The Lagrangian for the $(u, p_u)$ system and that of the $(y, p_y)$ are respectively, 
\begin{align}
 &L(u,u') 
 = \frac{1}{2\tilde{\alpha}} \frac{(u')^2}{u} + \frac{1}{2\tilde{\alpha}} u^2 -\frac{1}{2\tilde{\alpha}} su + \frac{M^2}{2\tilde{\alpha} u}, \\
 &\hat{L}(y,y')
 = -\frac{1}{2} \tilde{\alpha} \xi^2 (y')^2 -\frac{\tilde{\alpha}^2 \xi^3}{2} y' \left( y^2 + \frac{s}{\tilde{\alpha}^2 \xi^2} \right) 
 -\frac{1}{8} \tilde{\alpha}^3\xi^4 \left(y^2+ \frac{s}{\tilde{\alpha}^2\xi^2} \right)^2 + \tilde{\alpha} \eta M y.
\end{align}

%%%%%%%%%%%%%%%%%%%%%%%%%%%%%%%%%%%%%%%%%%%%%%%%%%%%%
\subsection{Non-perturbative function to $y$ asymptotic series}
%%%%%%%%%%%%%%%%%%%%%%%%%%%%%%%%%%%%%%%%%%%%%%%%%%%%%
In eq. \eqref{eom:y} for $y$ variable, the asymptotic expansion goes like 
\begin{equation}
 y_{\rm asym} = \frac{1}{|\tilde{\alpha} \xi|} \sqrt{-s} + \cdots.
\end{equation}
This is an asymptotic series distinct from $u_{\rm asym}$ of eq. \eqref{u:asym}
that begins with the planar solution of our original problem, and probes another corner of 
 the solution space of PII.
The way in which the non-perturbative  piece is investigated is completely analogues to
 that in sec \ref{npf:u}. 
Let $y$ and $\tilde{y}$ be two solutions with the same asymptotics, 
\begin{equation}
 z \equiv \tilde{y} - y,~~~~~
 |z| \ll |y|,~ |\tilde{y}|, 
\end{equation}
and we linearize eq. \eqref{eom:y} with respect to $z$. 
We just give our final result here: 
\begin{equation}
 z \approx (\text{const}) (-s)^{-\frac{1}{4}} \exp \left\{
 -\frac{2}{3|\tilde{\alpha}\xi|} (-s)^{\frac{3}{2}} 
 \right\}.
\end{equation}

%%%%%%%%%%%%%%%%%%%%%%%%%%%%%%%%%%%%%%%%%%%%%%%%%%%%%
%%%%%%%%%%%%%%%%%%%%%%%%%%%%%%%%%%%%%%%%%%%%%%%%%%%%%
\section{Direct evaluation of the exponential factor}
%%%%%%%%%%%%%%%%%%%%%%%%%%%%%%%%%%%%%%%%%%%%%%%%%%%%%
%%%%%%%%%%%%%%%%%%%%%%%%%%%%%%%%%%%%%%%%%%%%%%%%%%%%%
Let us turn our attention back to the original unitary matrix model and study work
done by a single eigenvalue as is stated in the introduction.
The computation of this object is given through that of the response of 
 the finite $N$ partition function $Z_N$ under  a move of the single eigenvalue in the presence of
 the remainder of $N-1$ eigenvalues lying inside the cut. 
 
Let $Z_{N-1, \mathrm{1\,max}}$ be the partition function 
 under the constraint that the single eigenvalue lies on the maximum of the effective potential 
 made by the remainder of the $N-1$ eigenvalues.
The object of our concern denoted by $S_{N,\,{\rm s.e.v.}}$ is extracted as 
\begin{equation}\label{I_N}
I_N \equiv \frac{Z_{N-1, \mathrm{1\,max}}}{Z_N} \equiv \ex^{-S_N,\,{\rm s.e.v.}}. 
\end{equation}
The square of eq. \eqref{I_N} may be regarded as a tunneling amplitude. 

To evaluate this in the large $N$ limit, we find it more expedient to convert the original
distribution on the unit circle into the one on the real axis, namely, to work actually
on a Hermitian matrix model . This goes as follows.

Recall that the finite $N$ partition function of our original unitary matrix model is
\begin{equation}
Z_{U(N)} = \frac{1}{\mathrm{Vol}(N)} \int [\de U] \exp\Big(
 \mathrm{Tr}\, W_U(U) 
\Big), 
\end{equation}
and
\begin{equation}
 (-W_U)(z) = - \frac{1}{2\underline{g_s}} \left(
  z + \frac{1}{z}  \right) 
  - M \log z.
\end{equation}
 (Here, we have made a change of variable $z \to -z$ for our convenience.)
The conversion into a Hermitian matrix model is simply given by the following $SL(2,\mathbf{C})$ transformation:
\begin{equation}
z = \frac{\im -w}{\im + w},~~~~~
z = \ex^{\im \theta}, ~~~ w\in \mathbf{R},~~~ w = \tan \frac{\theta}{2}. 
\end{equation}
The resultant Hermitian matrix model takes the form
\begin{equation}
 (-W)(w) = -\frac{2(1-w^2)}{1 + w^2} - 2M \ul{g_s} \log \frac{\im -w}{\im + w} + 2 \ul{g_s} N \log (1 + w^2).
\end{equation}

Let us evaluate $I_N$ in the planar limit 
\begin{align} 
 &I \equiv \lim I_N \equiv \ex^{- S_{\rm s.e.v.}}, \label{I} \\
 & S_{{\rm s.e.v.}}  = N \int_{\text{end point}}^{\infty,\,\rm maximum} \de x' f(x') ,  \label{S_sev:def}\\
 & f(x') = \frac{1}{2\tilde{S}} (-W)'(x') - 2 \mathrm{Re}\, \omega_{\rm pla}(x') . \label{f:def}
\end{align}
Here $\omega_{\rm pla}(x')$ is the resolvent  in the planar limit:
\begin{equation} 
 \omega_{\rm pla}(x') \equiv \lim_{N \to \infty} \frac{1}{N} \left\langle\!\!\!\left\langle \sum_{j=1}^N \frac{1}{x' - \lambda_j} \right\rangle\!\!\!\right\rangle . 
\end{equation}
In deriving eqs. \eqref{S_sev:def}, \eqref{f:def} we have used, for the hermitian matrix $H$, 
\begin{equation}
 \left. \log \det (x'-H)^2\right|_{\rm end\,point}^{\infty} 
 = 2 \int_{\rm end\,point}^{\infty} \de x' \,\mathrm{tr} \left( \frac{1}{x'-H}\right), 
\end{equation}
and ignored the size of the matrix by one. 

The evaluation of the planar resolvent is straightforward \cite{BIPZ,Mizo0411} and  we obtain
\begin{equation}
 \omega_{\rm pla}(x') = \frac{1}{4 \tilde{S}} \left(
  (-W)'(x') - \frac{8 \sqrt{x'^2 - b^2}}{\sqrt{1+b^2} (1+x'^2)^2}
 \right), ~~~~~
 b = \sqrt{\frac{\tilde{S}}{1-\tilde{S}}}, 
\end{equation}
and  eq. \eqref{f:def} is given by 
\begin{equation}\label{f}
 f(x') = \frac{4}{\tilde{S}} \frac{\sqrt{x'^2 - b^2}}{\sqrt{1+b^2} (1+x'^2)^2}, 
\end{equation}
and 
\begin{equation}\label{S_sev}
 S_{\rm s.e.v.} = \frac{4 N}{\tilde{S}} \int_b^{\infty} \de x' \frac{\sqrt{x'^2 - b^2}}{\sqrt{1+b^2} (1+x'^2)^2}. 
\end{equation}
Eq.\eqref{f} up to a factor $2\pi$ coincides with the density of the eigenvalues extended to the forbidden zone, namely continued outside to the cut. 
 In fact, the eigenvalue density one-form agrees with that of the original Gross-Witten:
 \begin{equation}
 \frac{1}{2\pi} |f(w)| \de w = \rho_{\rm GW}(\theta) \de\theta,~~~~~
 \rho_{\rm GW}(\theta) = \frac{1}{\pi \tilde{S}} \cos \frac{\theta}{2} \sqrt{\tilde{S} - \sin^2 \frac{\theta}{2}}. 
 \end{equation}

Let us carry out the actual computation of the integral \eqref{S_sev} in the continuum limit.
Let our scaling ansatz be $\tilde{S} = 1 - \beta a^2 t$, $a^3 = \frac{1}{N}$ and accordingly we scale $x'$ as $x'=\frac{1}{\sqrt{\beta}} a^{-1} \zeta $.
We obtain
 \begin{equation}
 S_{\text{s.e.v.}} = \frac{4}{3} (\beta t)^{\frac{3}{2}}. 
 \end{equation}
 The apparent parameter dependence gets resolved by recalling the renormalized(dressed)
 string coupling \cite{IOYano1909,IOYano2103}
\begin{equation}
 \kappa \equiv \frac{1}{N} \frac{1}{(1-\tilde{S})^{\frac{1}{2} (2 - \gamma_{st})}} 
 = (\beta t)^{-\frac{3}{2}}, 
\end{equation}
 and we obtain\footnote{For similar consideration and calculation in 2d noncritical string theory, see \cite{MV0304,Mart0305,MTV0305,AKK0306,HHIK0405}.} 
\begin{equation}\label{S}
 S_{\rm s.e.v.} = \frac{4}{3} \frac{1}{\kappa} . 
\end{equation}

%%%%%%%%%%%%%%%%%%%%%%%%%%%%%%%%%%%%%%%%%%%%%%%%%%%%%
%%%%%%%%%%%%%%%%%%%%%%%%%%%%%%%%%%%%%%%%%%%%%%%%%%%%%
\section{Comparison} 
%%%%%%%%%%%%%%%%%%%%%%%%%%%%%%%%%%%%%%%%%%%%%%%%%%%%%
%%%%%%%%%%%%%%%%%%%%%%%%%%%%%%%%%%%%%%%%%%%%%%%%%%%%%
Let us now compare the result eq. \eqref{ulz:general} in section three and that in section four, 
eqs. \eqref{I_N}, \eqref{I}, \eqref{S}. 
As is stated in the introduction, we will evaluate $\ul{z}(s)$ at the end point $K$ of the free energy integral
 eq. \eqref{W}. 
We obtain 
\begin{equation}
 \ul{z}(s) = (\text{const})' \kappa^{-\frac{1}{6}} \left( \frac{s}{K} \right)^{\frac{1}{4}} 
  \exp\left( -\frac{4}{3} \frac{1}{\kappa} \left( \frac{s}{K} \right)^{\frac{3}{2}} \right),
\end{equation}
and 
\begin{equation}
 \ul{z}(K) = (\text{const})'  \kappa^{-\frac{1}{6}} \exp \left( -\frac{4}{3} \frac{1}{\kappa} \right). 
\end{equation}
The exponential factor agrees with the value of $S_{\rm s.e.v.}$ in \eqref{S}.

%%%%%%%%%%%%%%%%%%%%%%%%%%%%%%%%%%%%%%%%%%%%%%%%%%%%%
%%%%%%%%%%%%%%%%%%%%%%%%%%%%%%%%%%%%%%%%%%%%%%%%%%%%%
\section*{Acknowledgments}
We thank Asato Tsuchiya for helpful discussions and insightful remarks.
We also thank Kazunobu Maruyoshi and Jaewon Song for discussions of related issues on this subject at Shuzenji workshop. 
The work of H.I. is supported in part by JSPS KAKENHI (19K03828, 23K03393, 23K03394).
%%%%%%%%%%%%%%%%%%%%%%%%%%%%%%%%%%%%%%%%%%%%%%%%%%%%%
%%%%%%%%%%%%%%%%%%%%%%%%%%%%%%%%%%%%%%%%%%%%%%%%%%%%%

%%%%%%%%%%%%%%%%%%%%%%%%%%%%%%%%%%%%%%%%%%%%%%%%%%%%%
%%%%%%%%%%%%%%%%%%%%%%%%%%%%%%%%%%%%%%%%%%%%%%%%%%%%%
%\appendix

%%%%%%%%%%%%%%%%%%%%%%%%%%%%%%%%%%%%%%%%%%%%%%%%%%%%%
%\section{}

%%%%%%%%%%%%%%%%%%%%%%%%%%%%%%%%%%%%%%%%%
%\bibliographystyle{arxiv}
%\bibliography{Painleve}
%%%%%%%%%%%%%%%%%%%%%%%%%%%%%%%%%%%%%%%%%

%%%%%%%%%%%%%%%%%%%%%%%%%%%%%%%%%%%%%%%%%%%%%%%%%%%%%

%%%%%%%%%%%%%%%%%%%%%%%%%%%%%%%%%%%%%%%%%%%%%%%%%%%%%

%%%%%%%%%%%%%%%%%%%%%%%%%%%%%%%%%%%%%%%%%%%%%%%%%%%%%
%%%%%%%%%%%%%%%%%%%%%%%%%%%%%%%%%%%%%%%%%%%%%%%%%%%%%
\end{document}